\documentstyle[nato]{crckapb}

\begin{opening}
\title{Singularity Analysis Towards 
Nonintegrability \protect\\ of Nonhomogeneous 
    Nonlinear Lattices} 
\author{Ken UMENO} 
\institute{Frontier Research Program\\ 
The Institute of Physical and Chemical Research (RIKEN)\\
2-1 Hirosawa, Wako, Saitama 351-01, Japan}
 
\end{opening}
\runningtitle{THE CRCKAPB STYLE FILE}
\newcommand{\beq}{\begin{equation}}
\newcommand{\eeq}{\end{equation}}
\newcommand{\beqa}{\begin{eqnarray}}
\newcommand{\eeqa}{\end{eqnarray}}

\begin{document}
 
\begin{abstract}
We show non-integrability of the nonlinear lattice of Fermi-Pasta-Ulam type
via  singularity analysis of normal variational equations 
 of Lam\'e type.  
\end{abstract}
\section{From a Nonlinear Lattice to Lam\'e Equations}
\setcounter{equation}{0}

We  consider the following one-dimensional lattice:
\begin{equation}
\label{eq:general}
  H=\frac{1}{2}\sum_{i=1}^{n}p_{i}^{2}+\frac{1}{2}
  \sum_{i=1}^{n+1}\upsilon(q_{i-1}-q_{i}),  
\end{equation}
 where 
\beq
\label{eq:potential}
  \upsilon(X)=\frac{\mu_{2}}{2}X^{2}+
  \frac{\mu_{4}}{4}X^{4}+\cdots +\frac{\mu_{2m}}{2m}
  X^{2m}.
\eeq

 Fermi-Pasta-Ulam(FPU) lattice \cite{fpu} is a special type of the systems with the 
 potential function (\ref{eq:potential})  
  as follows:
\begin{equation}
\label{eq:ulam}
  H_{FPU}=\frac{1}{2}\sum_{i=1}^{n}p_{i}^{2}+\frac{\mu_{2}}{2}
  \sum_{i=1}^{n+1}(q_{i-1}-q_{i})^{2}+
  \frac{\mu_{4}}{4}\sum_{i=1}^{n+1}(q_{i-1}-q_{i})^{4}.
\end{equation}

If we impose the fixed boundary condition as 
\beq
\label{eq:fixbound}
  q_{0}=q_{n+1}=0, \quad n=\mbox{odd},
\eeq
it is easy to check that 
\beq
\label{eq:compsol}
\Gamma: \quad  q_{1}=C\phi(t),q_{2}=0,q_{3}=-C\phi(t),\cdots,q_{n-1}=0,
q_{n}=(-1)^{\frac{n-1}{2}}
 C\phi(t)
\eeq
is a special solution.
  Thus, the equation of \(\phi (t)\) is equivalent  to the 
  following Hamiltonian 
  system with one degree of freedom:
\begin{equation}
\label{eq:oneh}
  \ddot{\phi}+2\mu_{2}\phi+2\mu_{4} C^{2}\phi^{3}+\cdots+2\mu_{2m}C^{2m-2}
  \phi^{2m-1}=0,
\end{equation}
where Hamiltonian is   
\begin{equation}
\label{eq:determin}
   H(\phi,\dot{\phi})=
  \frac{1}{2}(\dot{\phi})^{2}+\mu_{2}\phi^{2}+\frac{\mu_{4} C^{2}}{2}\phi^{4}
  +\cdots +\frac{\mu_{2m} C^{2m-2}}{m}
  \phi^{2m}
  =\mbox{Const.}
\end{equation}
 Then the total energy \(\epsilon\) is given by  
\begin{equation}
\label{eq:energyr}
  \epsilon=H=
  H(\phi,\dot{\phi})\frac{n+1}{2}C^{2}=
  \frac{n+1}{2}C^{2}(\mu_{2}+\frac{1}{2}\mu_{4} C^{2}+\cdots\frac{1}{m}\mu_{2m}
  C^{2m-2})
\end{equation} 
for the initial condition (\ref{eq:compsol}).
In the case of the FPU lattice,  we can determine \(C\) as follows: 
\beq
  C=\sqrt{\frac{\sqrt{\mu^{2}_{2}+\frac{4\epsilon}{n+1}\mu_{4}}
  -\mu_{2}}
  {\mu_{4}}}.
\eeq
By combining (\ref{eq:determin}) with (\ref{eq:energyr}),
the underlying equation of \(\phi(t)\) can be rewritten by  
the
differential equation of \(\phi(t)\) as
\begin{equation}
\label{eq:finalde}
 \frac{1}{2}(\dot{\phi})^{2}=\gamma_{2}
 (1-\phi^{2})+\frac{\gamma_{4}}{2}(1-\phi^{4})
 +\cdots+\frac{\gamma_{2m}}{m}(1-\phi^{2m}),
\end{equation}
where 
\begin{equation}
  \gamma_{2m}(\epsilon,\{\mu_{2j}|j=1,\cdots,m\})\equiv \mu_{2m} C^{2m-2}.
\end{equation}

In the case of the FPU lattices (\ref{eq:ulam}), 
the solution of  
this differential equation (\ref{eq:finalde}) with the condition 
\beq
 \gamma_{2m=4}\ne 0
\eeq
 is given explicitly by the 
 elliptic function 
\begin{equation}
  \phi(t)=cn(k;\alpha t),
\end{equation}
 where
\begin{equation}
\label{eq:3rel1}
   \alpha=\sqrt{2\gamma_{2}+2\gamma_{4}}, 
   \quad k=\sqrt{\frac{\gamma_{4}}
   {2\gamma_{2}+2\gamma_{4}}},
\end{equation} 
\(cn(k;\alpha t)\) is the  Jacobi 
{\it cn} elliptic 
function, and \(k\) is  the modulus of the elliptic integral. 
We remark  that because  
\beq
 \gamma_{2}+\gamma_{4}=\mu_{2}+C^{2}\mu_{4}=
 \sqrt{\mu_{2}^{2}+\frac{4\epsilon}{n+1}\mu_{4}}>0,
\eeq
holds 
for \(\mu_{4}>0,\mu_{2}\geq 0\),   
the modulus of the elliptic function \(k\) satisfies the following relation:
\beq
  0\leq k \leq \frac{1}{\sqrt{2}}.
\eeq
Thus, the  special solutions
of the  FPU lattices  for \(\mu_{4}>0,\mu_{2}\geq 0\)
have the two {\it fundamental periods} in the complex 
time plane as follows: 
\begin{equation}
\label{eq:rel2}
    T_{1}(\epsilon,\mu)=\frac{2K(k)}{\alpha},
    \quad T_{2}(\epsilon,\mu)=\frac{2K(k)+2iK'(k)}{\alpha},
\end{equation}    
 where  \(K(k)\) and \(K'(k)\) are 
 the complete elliptic integrals of the first kind: 
\begin{equation}
   K(k)=\int_{0}^{1}\frac{dv}{\sqrt{(1-v^{2})(1-k^{2}v^{2})}},\quad
   K'(k)=\int_{0}^{1}\frac{dv}{\sqrt{(1-v^{2})(1-(1-k^{2})v^{2})}}.
\end{equation} 
 Poles are located  at 
 \(t=\tau,\) where 
 \(\tau=\frac{2K(k)}{\alpha}+i\frac{K'(k)}{\alpha}\quad (\mbox{mod }
 T_{1},T_{2})\)
 in the parallelogram of each period cell. 
 Let 
 us consider the variational equations along these special solutions. 
 The  variational equations are obtained by  
\begin{equation}
\label{eq:firstve}
\begin{array}{l}
  \dot{\eta_{j}}= 
  \ddot{\xi_{j}}=
  -\sum_{k=1}^{n}\left.\frac{\partial^{2}V}{\partial q_{k}\partial q_{j}}
  \right|_{\Gamma}
  \xi_{k}\\
  =
 -(\gamma_{2}+3\gamma_{4} \phi^{2}+5\gamma_{6}\phi^{4}+\cdots+
 (2m-1)\gamma_{2m}\phi^{2m-2})
 (2\xi_{j}-\xi_{j-1}-\xi_{j+1})
  \mbox{  for  }1\leq j\leq n,
\end{array}
\end{equation}
 where
   \(\xi_{0}=\xi_{n+1}=\eta_{0}=\eta_{n+1}=0\) and \(\xi_{j}=\delta q_{j},
   \eta_{j}=\delta p_{j}\quad(1\leq j\leq n)\).\\
 Moreover,  these linear 
  variational equations  in the  form of the vector 
\begin{equation}
\frac{d^{2}}{dt^{2}}
\mbox{\boldmath$\xi$}=-(\gamma_{2}+\cdots+(2m-1)\gamma_{2m}
\phi^{2m-2})
\left[
\begin{array}{ccccc}
2 & -1 & 0 & \cdots & 0 \\
-1 & 2 & -1 & \cdots & 0 \\
0 & -1 & 2 & -1 & \cdots \\
\cdots & \cdots & \cdots & \cdots & \cdots \\
0 & \cdots & 0 & -1 & 2  
\end{array}
\right]
\mbox{\boldmath$\xi$}
\end{equation}
can be decoupled as follows.
 After we note that  the eigenvalues of the \(n\times n\) symmetric matrix 
\begin{equation}
\mbox{\boldmath$G$}=
\left[
\begin{array}{ccccc}
2 & -1 & 0 & \cdots & 0 \\
-1 & 2 & -1 & \cdots & 0 \\
0 & -1 & 2 & -1 & \cdots \\
\cdots & \cdots & \cdots & \cdots & \cdots \\
0 & \cdots & 0 & -1 & 2  
\end{array}
\right]
\end{equation}
are obtained as \(\{4\mbox{sin}^{2}(\frac{j\pi}{2(n+1)})|1\leq 
  j\leq n\}\) 
 by a normal orthogonal transformation 
\(\mbox{\boldmath$G$}\rightarrow
\mbox{\boldmath$OGO^{-1}$}\), 
 the variational equations (\ref{eq:firstve}) are
 rewritten in the decoupled form:
\begin{equation}
\label{eq:cve}
  \ddot{\xi'_{j}}(t)=-4\mbox{sin}^{2}(\frac{j\pi}{2(n+1)})
  (\gamma_{2}+3\gamma_{4}\phi^{2}+\cdots+
  (2m-1)\gamma_{2m}\phi^{2m-2})\xi'_{j}(t)\quad(1\leq j \leq n),
\end{equation}
where \(\mbox{\boldmath$\xi'$}=\mbox{\boldmath$O\xi$}\).
Clearly, these equations are 
written in the form of vector {\it Hill's equation}\cite{3hill}
\beq
  \frac{d^{2}\mbox{\boldmath$\xi'$}}{dt^{2}}+\mbox{\boldmath$A$}(t)
  \mbox{\boldmath$\xi'$}=0,\quad
  \mbox{\boldmath$A$}(t+T)=\mbox{\boldmath$A$}(t),
\eeq
 where \(T=T_{1},T_{2}\) in the case of \(m=2\).  
For \(j=\frac{n+1}{2}\),  we have the relation  
\begin{equation}
\label{eq:mannaka}
\xi'_{\frac{n+1}{2}}=\sqrt{\frac{2}{n+1}}(\xi_{1}-\xi_{3}+\xi_{5}+\cdots+(-1)^{\frac{n-1}{2}}
\xi_{n}).
\end{equation}
Thus,  the corresponding variational equation 
\begin{equation}
\label{eq:tangenve}
  \ddot{\xi'}_{\frac{n+1}{2}}=-2(\gamma_{2}+3\gamma_{4}\phi^{2}+\cdots
  +(2m-1)\gamma_{2m}\phi^{2m-2})\xi'_{\frac{n+1}{2}}(t)
\end{equation} 
has a time-dependent integral 
\(I(\mbox{\boldmath$\xi$},\dot{\mbox{\boldmath$\xi$}};t)
\equiv I(\mbox{\boldmath$\xi$},\mbox{\boldmath$\eta$};t)\)
because  
\beq
\begin{array}{l}
  I(\mbox{\boldmath$\xi$},\mbox{\boldmath$\eta$};t)=
   \mbox{\boldmath$D$}H\equiv 
  (\mbox{\boldmath$\eta$}\cdot
  \frac{\partial}{\partial\mbox{\boldmath$p$}}+\mbox{\boldmath$\xi$}\cdot
  \frac{\partial}{\partial\mbox{\boldmath$q$}})H
  =\mbox{\boldmath$\eta$}\cdot\mbox{\boldmath$p$}+\mbox{\boldmath$\xi$}
  \cdot\mbox{\boldmath$V_{q}$}\\
  =C\dot{\phi}(\eta_{1}-\eta_{3}+\eta_{5}+
  \cdots+(-1)^{\frac{n-1}{2}}\eta_{n})\\
  +2(C\gamma_{2}\phi+C\gamma_{4}\phi^{3}+\cdots+C\gamma_{2m}\phi^{2m-1}
  )(\xi_{1}-\xi_{3}+\xi_{5}+\cdots+(-1)^{\frac{n-1}{2}}
  \xi_{n}),
\end{array}
\eeq
where 
\beq
\begin{array}{l}
\frac{1}{C}\frac{dI}{dt}=\dot{\phi}(\ddot{\xi_{1}}-\ddot{\xi_{3}}+
\cdots+(-1)^{\frac{n-1}{2}}
\ddot{\xi_{n}})\\
+2\dot{\phi}(\gamma_{2}+3\gamma_{4}\phi^{2}+\cdots +
(2m-1)\gamma_{2m}\phi^{2m-2})
(\xi_{1}-\xi_{3}+\cdots+(-1)^{\frac{n-1}{2}}
\xi_{n})
=0. 
\end{array}
\eeq
We call Eq. (\ref{eq:tangenve})  the {\it  tangential variational 
equation}.
On the other hands, a \((2n-2)\)-dimensional 
{\it normal variational equation}(NVE) is given  by 
the equation of (\ref{eq:cve}) 
with the tangential variational equation (\ref{eq:tangenve}) 
removed as follows:
\begin{equation}
\label{eq:nveev}
\begin{array}{l}
  \dot{\eta'}_{j}=
  -4\mbox{sin}^{2}(\frac{j\pi}{2(n+1)})(\gamma_{2}+3\gamma_{4} \phi^{2}+\cdots
  +(2m-1)\gamma_{2m}\phi^{2m-2})
  \xi'_{j},\\
  \dot{\xi'}_{j}=\eta'_{j}\quad\mbox{for }1\leq j(\ne \frac{n+1}{2}) 
  \leq n.
\end{array}
\eeq
  In case of the FPU lattice,  
the normal variational equation (\ref{eq:nveev}) 
 becomes 
 the {\it Lam\'e 
 equation} \cite{whit}
\begin{equation}
 \frac{d^{2} y}{dt^{2}}-(E_{1}sn^{2}(k;\alpha t)+E_{2})y=0,
\end{equation}
 where \(E_{1}=12\frac{1}{\alpha^{2}k^{2}}\mbox{sin}^{2}(\frac{j\pi}{2(n+1)})
 \) and \(E_{2}\) are constants.

\section{Non-integrability Theorem}

   Morales and Sim\'o obtained 
 the following theorem on the non-integrability based on 
 the application of Picard-Vessiot theory to Ziglin's analysis\cite{zig1,zig2}
 for Hamiltonian systems 
with two degrees of freedom.
\newtheorem{th3}{Theorem}
\begin{th3}[Morales and Sim\'o \cite{mor},1994]
\label{th3:fputheorem1}

When the normal reduced variational equation is of Lam\'e type, if 
\(A\equiv E_{1}\alpha^{2}k^{2}\ne m(m+1), m\in \mbox{\boldmath$N$}\)
 and the 
Lam\'e equation satisfying this condition on \(A\) is not algebraically 
solvable(Brioschi-Halphen-Crawford and Baldassarri solutions), then the 
the initial Hamiltonian system does not have a first integral, meromorphic 
in a connected neighborhood of the integral curve \(\Gamma\), which is 
functionally independent together with \(H\). 
\end{th3}

In case of the present  analysis, 
\(A\) is given by the following formula:
\beq
  A=E_{1}\alpha^{2}k^{2}=12\mbox{sin}^{2}(\frac{j\pi}{2(n+1)})
    =6(1-\mbox{cos}(\frac{j\pi}{n+1})).
\eeq
We can easily check that  
\(\mbox{cos}(\frac{j\pi}{n+1}) \notin \mbox{\boldmath$Q$} \) if and   
only if \(j\notin \{\frac{n+1}{3},\frac{n+1}{2},\frac{2(n+1)}{3}\}\).
When \(A\notin \mbox{\boldmath$Q$}\), the above condition on the 
algebraic solvability of the Lam\'e equation is not satisfied.  
Thus, to check the algebraic solvability of the Lam\'e equations 
\beq
\label{eq:alame}
\frac{d^{2} \xi_{j}}{dt^{2}}-(
\frac{12}{\alpha^{2} k^{2}}\mbox{sin}^{2}(\frac{j\pi}{2(n+1)})
sn^{2}(k;\alpha t)+E_{2})\xi_{j}=0\quad (j\ne \frac{n+1}{2}), 
\eeq
it is sufficient  to  examine the following two cases:
\beq
   A=6(1-\mbox{cos}(\frac{1\pi}{3}))=3, \quad
   A=6(1-\mbox{cos}(\frac{2\pi}{3}))=9.
\eeq
It is known \cite{bald} that   
the condition on \(A\) for the Brioschi-Halphen-Crawford solutions 
is given by 
\beq
  A=m(m+1),\quad m+\frac{1}{2}\in \mbox{\boldmath$N$},  
\eeq 
 and that the condition on \(A\) for the Baldassarri solutions 
 is given by 
\beq
   A=m(m+1), \quad m+\frac{1}{2}\in \frac{1}{3}\mbox{\boldmath$Z$}
  \cup \frac{1}{4}\mbox{\boldmath$Z$}\cup \frac{1}{5}\mbox{\boldmath$Z$}
  \setminus \mbox{\boldmath$Z$}.
\eeq
  However, the following relations
\beq
m(m+1)=3 \rightarrow m=\frac{-1\pm\sqrt{13}}{2}\notin \mbox{\boldmath$Q$},
\quad
  m(m+1)=9 \rightarrow m=\frac{-1\pm\sqrt{37}}{2}\notin \mbox{\boldmath$Q$}
\eeq    
hold,  which  guarantee that {\it all} \(n-1\)  
Lam\'e  equations (\ref{eq:alame}) 
do not belong to the solvable 
case. In case of the systems with \(n\) degrees of freedom, we have 
\(n-1\) Lam\'e equations which corresponds to \(n-1\) 
normal variational equations.

Thus, according to the steps in Ref. \cite{mor} we obtain the following 
theorem:
\begin{th3}
\label{th3:fputheorem}
The FPU lattice for \(\mu_{4}>0, \mu_{2}\geq 0\) does 
not have \(n-1\) first integrals, meromorphic in a 
connected neighbourhood of the integral curve \(\Gamma\), which are 
functionally independent together with \(H\).
\end{th3}

We remark here that this theorem on the non-integrability 
does not depend on the total energy  in contrast with the result 
about the non-integrability proof of the FPU lattice in the low energy 
limit \cite{ku2} based on  non-resonance checking \cite{4umth} and 
the result about the non-integrability of the FPU lattice in the 
high energy limit based on the Kowalevski exponents of the 
homogeneous systems \cite{yo}. Here, it is conjectured that more general 
nonhomogeneous 
nonlinear lattice (\ref{eq:general}) would be also non-integrable 
in the sense of the present analysis.

%cccccccccccccccccccccccccccccccccccccccccccccccccccccccccccccccccccc
\section*{Acknowledgements}

The present author would like to thank Prof. S. Dovysh, 
 Prof. J. J. Morales, Prof. J. Moser, and Prof. H. Yoshida for valuable 
discussions. He is grateful to Prof. C. Sim\'o, Prof. A. Delshams, 
Prof. de la Llave and Prof. T. Konishi for their kind hospitalitiy 
during the 3DHAM95 meeting. He thanks Prof. S. Amari for his continual 
 encouragement. He appreciates    
  support from
 the Special Researchers Program of Basic 
  Science at the 
 Institute of Physical and Chemical Research (RIKEN) and from the 
 Program of the Complex Systems Workshop  at the International 
 Institute for Advanced Study (IIAS).

\end{document}